# Formal Concept Analysis Based Association Rules Extraction

Ben Boubaker Saidi Ourida[1] and Tebourski Wafa[2]

[1] Computer Science Department, High Institute of Management, University of Tunis,
Bouchoucha, Bardo-Tunis, Tunisie

[2] Computer Science Department, High Institute of Management, University of Tunis,
Bouchoucha, Bardo-Tunis, Tunisie

## ABSTRACT

*Generating a huge number of association rules reduces their utility in the decision making process, done by domain experts. In this context, based on the theory of Formal Concept Analysis, we propose to extend the notion of Formal Concept through the generalization of the notion of itemset in order to consider the itemset as an intent, its support as the cardinality of the extent and its relevance which is related to the confidence of rule. Accordingly, we propose a new approach to extract interesting itemsets through the concept coverage. This approach uses a new quality-criteria of a rule: the relevance bringing a semantic added value to formal concept analysis approach to discover association rules.*

## 1. INTRODUCTION

Given the information density and mass accumulation of data, it was crucial to explore this information in order to extract meaningful knowledge.

The "Concept" is a couple of intent and extent aiming to represent nuggets of knowledge. Recently, researchers have been striving to build theoretical foundations for data-Mining based on Formal Concept Analysis [1,2,3]. Several interesting proposals have appeared, related to association rules [4].

In this paper, we introduce a novel approach of association rules mining based on Formal Concept Analysis.

The remainder of the paper is organized as follows. We outline in Section 2 the association rules derivation problem. Section 3 introduces the mathematical background of FCA and its connection with the derivation of association rule bases. We present, in Section 4, an heuristic algorithm to calculate the optimal itemsets from rows of data. Section 5 describes the results of the experimental study. Illustrative examples are given throughout the paper. Section 6 concludes this paper and points out future research directions.

## 2. ASSOCIATION RULES DERIVATION

Commonly, the number of the generated association rules grows exponentially with the number of data rows and attributes. This can reach hundreds' of thousands using only some thousands of data rows. So, their comprehension and their interpretation become a hard task.
To remedy to this problem, several methods were proposed [17].
The commonly generated thousands and even millions of rules – among which many are redundant (Bastide et al., 2000; Stumme et al., 2001; Zaki, 2004) –[5, 6, 7, 8] encouraged the proposal of more discriminating techniques to reduce the number of reported rules.
This pruning can be based on patterns defined by the user (user-defined templates), on Boolean operators (Meo et al., 1996; Ng et al., 1998; Ohsaki et al., 2004; Srikant et al., 1997) [9,10,11,12].

The number of rules can be greatly reduced through pruning focusing on additional information namely the taxonomy of items (Han, & Fu, 1995) or on a metric of specific interest (Brin et al., 1997) (e.g., Pearson's correlation or $\chi^2$-test) [13, 14]. More advanced techniques that produce only lossless information limited number of the entire set of rules, called generic bases (Bastide et al., 2000). The generation of such generic bases heavily draws on a battery of results provided by formal concept analysis (FCA) (Ganter & Wille, 1999) [15].

Primitively, the pruning strategy of association rules is based on crucial techniques namely the frequency of the generated pattern through discarding all the itemsets having a support less than MinSup, and the strength of the dependency between premise and conclusion by pruning all the rules having a confidence less than MinConf.

To prune effectively the extracted association rules, some authors [16] introduce another measures. In fact, Bayardo et al propose the conviction measure. Moreover, Cherfi et al [17] suggest five different measures such as the benefit (interest) and the satisfaction. Maddouri et al provide the gain measure [18].

In this paper, we introduce a new measure: the *relevance*.

Indeed, it is backboned on the Formal Concept Analysis [19, 20]. Assuming that an itemset is completely represented by a formal concept as a couple of intent (the classic itemset) and extent (its support), it combines the support of the rule with the length of the itemset. Thus, we propose to include a semantic aspect on association rules extraction by taking into account the confidence measure during the selection of frequent itemsets during association rules generation.

## 3. MATHEMATICAL BACKGROUND

We recall some crucial results inspired from the Galois lattice-based paradigm in FCA and its interesting applications to association rules extraction.

### 3.1. Preliminary notions

In the remainder of the paper, we use the theoretical framework presented in [20].

Let O be a set of objects, P a set of properties and R a binary relation defined between O and P [19, 20].

TABLE 1. FORMAL CONTEXT

| O\I | A | B | C | D |
|---|---|---|---|---|
| o1 | 1 | 1 | 0 | 0 |
| o2 | 1 | 1 | 0 | 0 |
| o3 | 0 | 1 | 1 | 0 |
| o4 | 0 | 1 | 1 | 1 |
| o5 | 0 | 0 | 1 | 1 |

**Definition 1** [19]: A formal context (O, P, R) consists of two sets O and P and a relation R between O and P. The elements of O are called the objects and the elements of P are called the properties of the context. In order to express that an object o is in a relation R with a property p, we write oRp or (o, p)∈R and read it as "the object o has the property p".

**Definition 2** [19]: For a set A⊆O of objects and a set B⊆P of properties, we define :

The set of properties common to the objects in A :

A▶={p∈P | oRp for all o∈A}

The set of objects which have all properties in B :

B◀={o∈O | oRp for all p∈B}

The couple of operators (▶, ◀) is a Galois Connection.

**Definition 3** [19]: A formal concept of the context (O, P, R) is a pair (A, B) with A⊆O, B⊆P, A▶=B and B◀=A.
We call A the extent and B the intent of the concept (A, B).

**Definition 4** [19]: The set of all concepts of the context (O, P, R) is denoted by Φ (O, P, R). An ordering relation (<<) is easily defined on this set of concepts by :
(A1, B1) << (A2, B2) ⇔ A1⊆A2 ⇔ B2⊆B1.

FIGURE 1. CONCEPT LATTICE OF THE CONTEXT (O, P, R)

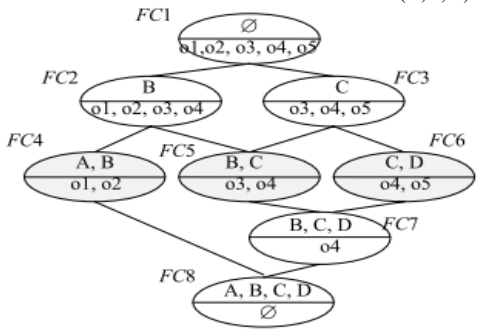

In this subsection, we remind basic theorem for Concept Lattices [19]:

Φ (O, P, R, <<) is a complete lattice. It is called the *concept lattice* or *Galois lattice* of (O, P, R), for which infimum and supremum can be described as follow:

$Sup_{i \in I} (A_i, B_i) = ((\cup_{i \in I} A_i)▶◀, (\cap_{i \in I} B_i))$

$Inf_{i \in I} (A_i, B_i) = (\cap_{i \in I} A_i, (\cup_{i \in I} B_i)◀▶)$

**Example** [18]: table 1 illustrates the notion of formal context (O, P, R). The latter is composed of five objects {o1, o2, o3, o4, o5} and four properties {A, B, C, D}. The concept lattice of this context is drawn in Figure 1 containing eight formal concepts.

**Definition 5** [19]: Let (o, p) be a couple in the context (O, P, R). The pseudo-concept PC containing the couple (o, p) is the union of all the formal concepts containing (o,p).

**Definition 6** [20]: A coverage of a context (O, P, R) is defined as a set of formal concepts CV={RE$_1$, RE$_2$, ..., RE$_n$} in Φ (O, P, R), such that any couple (o, p) in the context (O, P, R) is included in at least one concept of CV.

FIGURE 2. ILLUSTRATIVE EXAMPLE OF PSEUDO-CONCEPT, OPTIMAL CONCEPT, AND NON OPTIMAL CONCEPT CONTAINING THE COUPLE (o3,B).

| O \ I | A | B | C |
|---|---|---|---|
| o1 | 1 | 1 | 0 |
| o2 | 1 | 1 | 0 |
| o3 | 0 | 1 | 1 |

a. Pseudo-concept of (o3,B)

| O \ I | A | B | C |
|---|---|---|---|
| o1 | 1 | 1 | 0 |
| o2 | 1 | 1 | 0 |
| o3 | 0 | 1 | 1 |

b. Optimal concept of (o3,B)

| O \ I | A | B | C |
|---|---|---|---|
| o1 | 1 | 1 | 0 |
| o2 | 1 | 1 | 0 |
| o3 | 0 | 1 | 1 |

c. Non optimal concept of (o3,B)

**Example** [18]:
Considering the formal context (O, P, R) depicted by table 1, the figure 2.a represents the pseudo-concept containing the couple (o3, B) being the union of the concepts FC$_2$ and FC$_5$.

A coverage of the context is formed by the three concepts: {FC$_4$, FC$_5$, FC$_6$} such as:
- FC$_4$ is the concept containing the items ({o1, o2}, {A, B});
- FC$_5$ is the concept containing the items ({o3, o4}, {B, C});
- FC$_6$ is the concept containing the items ({o4, o5}, {C, D}).

The lattice constitutes concept coverage.

493

## 4. DISCOVERY OF OPTIMAL ITEM-SETS

The most expensive step to derive association rules is the computation of the frequent itemsets [4]. Indeed, this step consists of applying, iteratively, some heuristics to calculate candidate itemsets. At the iteration i, we combine the itemsets of the iteration i-1. After that, the support threshold (MinSup) is used to prune non-frequent candidates. The itemsets of iteration i-1, are also discarded. We keep the remaining itemsets of the latest iteration n with n is the number of properties in the formal context.

Two characteristics are used in the association rules derivation:

- Support: is the cardinality of the set of objects which verify the rule. In Formal Concept Analysis, it refers to the extent of a formal concept.
- Cardinality of the Itemset: is the number of properties of the itemset. In Formal Concept Analysis, it refers to the intent of a formal concept.

Assuming that the intent is not sufficient to represent the association rule, the latter is completely related to a formal concept namely both its intent and its extent. Having a support represented by the cardinality of the extent, a highly qualified selection of itemsets must be done according to the intent of the formal concept of the rule. Moreover, the association rule generated from the formal concept should take into account the quality of the relationship between the head and the body of the rule.

To formalize the new criterions, we give the following definitions.

**Definition 7** : Let $FC_i = (A_i, B_i)$ be a formal concept. We define:

- Length of a concept $FC_i$: the number of properties in the intent $B_i$ of the concept.
- Width of a concept $FC_i$: the number of objects in the extent $A_i$ of the concept.
- Conf of a concept $FC_i$: the maximum confidence of the set of rules generated from the concept $FC_i$.
- Relevance of a concept: is a function of the width the length and the confidence of the concept, given by:

$$Relevance(FC_i) = (length(FC_i) + conf(FC_i)) * (length(FC_i) + width(FC_i))$$

The relevance measure depends on the number of properties. In fact, a less number of properties, a less value of relevance is noted. Having more properties induced to higher relevance. Moreover, if we have a higher number of properties and objects, a higher value of relevance is affected. Increasing the confidence of the concept induce to higher value of relevance.

**Definition 8** : A formal concept $FC_i = (A_i, B_i)$ containing a couple (o, p) is said to be optimal if it maximizes the *relevance* function.

**Definition 9** [20]: A coverage CV={ $FC_1$, $FC_2$, ..., $FC_k$ } of a context (O, P, R) is optimal if it is composed by optimal concepts.

**Example** [18]: An illustrative example of the pseudo-concept is sketched by figure 2. b represents the optimal concept $FC_5$ containing the couple (o3, B). Figure 2.c represents the non optimal concept $FC_2$ containing the couple (o3, B).

The optimal coverage of the context (O, P, R) is formed by three optimal concepts: {$FC_4$, $FC_5$, $FC_6$}. $FC_4$ is the concept containing the items ({o1, o2}, {A, B}). $FC_5$ is the concept containing the items ({o3, o4}, {B, C}). $FC_6$ is the concept containing the items ({o4, o5}, {C, D}).

494

## 4.1 Heuristic Searching for Optimal Concept

The pseudo-concept, denoted by PCF, containing the couple (o, p), is the union of all the concepts containing (o, p). It is computed according to the relation R by the set of objects described by p, then {p}▶, and the set of properties describing the object o, so {o}◀. Where (▶,◀) is the Galois connection of the context (O, P, R).

When we determinate the pseudo-concept PCF, two cases are considered:
- *Case 1*: PCF forms a formal concept.

If no zero is found in the relation/matrix representing PCF, then, PCF is the optimal concept. So, the algorithm stops.
- *Case 2*: PCF is not a formal concept.

If some zero entries are found in the relation/matrix representing PCF, we will look for more restraint pseudo-concepts within the pseudo-concept PCF.

So, we consider the pseudo-concepts containing the couples like (X, p) or (o, Y). These concepts contain, certainly, the couple (o, p).

***The considered heuristic is the optimal concept is certainly included in the optimal pseudo-concept***.

So, we should generate all possible rules from the pseudo-concepts containing the couples like (X, p) or (o, Y). Then we compute the corresponding confidences and choose the maximum value between them. After that, we calculate the relevance value. Finally, we choose the pseudo-concept having the greatest value of the Relevance function to be the new PCF.

This heuristic procedure (of case 2) is repeated until PCF becomes a formal concept. To calculate the relevance of a pseudo-concept, we introduce the general form of the previous function:

Definition 10: Let $PCF_i = (A_i, B_i, R_i)$ be a pseudo-concept, where $R_i$ is the restriction of the binary relation R, to the subsets $A_i$ and $B_i$. We define the:
- Length of a pseudo-concept $PCF_i$: the number of properties in $B_i$.
- Width of a pseudo-concept $PCF_i$: the number of objects in $A_i$.
- Confidence of a pseudo-concept $PCF_i$: is the maximum of confidence found when we generate the set of rules extracted from the pseudo-concept $PCF_i$.
- Size of a pseudo-concept $PCF_i$: the number of couples (of values equal to 1) in the pseudo-concept. When $PCF_i$ is a formal concept, we have:

$$Size(PCF) = (length(PCF) * width(PCF))$$

- Relevance of a pseudo-concept is a function of the width, the length, the size and the confidence given by:

$$Relevance(PCF) = \left[\frac{Size(PCF)}{length(PCF) + conf(PCF)}\right] * [(length(PCF) + width(PCF)) - Size(PCF)]$$

## 4.2 Algorithm for Optimal Coverage

The problem of covering a binary relation by a set of optimal concepts is expressed through covering a binary matrix by a number of its complete sub-matrix. The latter is is a matrix having all its entries equal to '1'. This problem, being NP-Complete problem, has been the subject of several previous works. However, we found out necessary the proposition of an approximate polynomial algorithm called *Semantic Based on Formal Concept Analysis Approach SFC2A*.

Let R be the binary relation to cover. The proposed solution is to divide R into n packages (subsets): $P_1, ..., P_n$.

Each package symbolizes one or more couples.

The key idea of SFC2A algorithm is to build incrementally the optimal coverage of R:

*(i)* The first step, covering the relation $R_1 = P_1$ by $CV_1$.

*(ii)* The $i^{th}$ step, let $R_{i-1} = P_1 \ldots P_{i-1}$ and let $CV_{i-1}$ be its optimal coverage. Building the optimal coverage $CV_i$ of $R_i = R_{i-1} \cup P_i$ using the initial coverage $CV_{i-1}$ and the package $P_i$.

*(iii)* The $n^{th}$ step, finally, finding a set of concepts covering the relation R.

*Algorithm SFC2A*
*Begin*
*Let R be partitioned to n packages $P_1, \ldots, P_n$.*
*Let $CV_0 := \emptyset$.*
*FOR i=1 to n DO*
    *Sort the couples of $P_i$ by the pertinence of their pseudo-concepts*
    *While $(P_i \neq \emptyset)$ Do*
- *Select a couple (a, b) in $P_i$ by the sorted order of the relevance function*
- *Search PC : the pseudo-concept containing (a, b) within $R_i = CV_{i-1} \cup P_i$*
- *Search FC: the optimal concept containing (a,b) within PC*

*$CV_i := (CV_{i-1} - \{r \in CV_{i-1} / r \subseteq FC\}) \cup \{FC\}$: Delete all the redundant concepts from $CV_i$*

        $P_i := P_i - \{(X,Y) \in P_i / (X,Y) \in FC\}$
    *End While*
*End FOR*
*End*.

| P\O | A | B | C | D |
|---|---|---|---|---|
| o1 | 1 | 1 | 0 | 0 |
| o2 | 1 | 1 | 0 | 0 |
| o3 | 0 | 1 | 1 | 0 |

a. Optimal coverage of the context ({o1,o2,o3}{A,B,C,D})

| P\O | A | B | C | D |
|---|---|---|---|---|
| o1 | 1 | 1 | 0 | 0 |
| o2 | 1 | 1 | 0 | 0 |
| o3 | 0 | 1 | 1 | 0 |
| o4 | 0 | 1 | 1 | 1 |

b. Case 1 : Coverage of the context ({o1,o2,o3,o4}{A,B,C,D})

| P\O | A | B | C | D |
|---|---|---|---|---|
| o1 | 1 | 1 | 0 | 0 |
| o2 | 1 | 1 | 0 | 0 |
| o3 | 0 | 1 | 1 | 0 |
| o4 | 0 | 1 | 1 | 1 |

c. Case 2 : Coverage of the context ({o1,o2,o3,o4}{A,B,C,D})

FIGURE 3. INCREMENTATION STEP WHEN ADDING P4

Example: Let R be the relation to cover as highlighted by table 1. R is partitioned into five packages:
$P_1 = \{o1\} \times \{A, B\}$,
$P_2 = \{o2\} \times \{A, B\}$,
$P_3 = \{o3\} \times \{B, C\}$,
$P_4 = \{o4\} \times \{B, C, D\}$ and
$P_5 = \{o5\} \times \{C, D\}$.
Initially, R is an empty relation and in each step we add a package.

Figure 3 presents the incrementation step when adding $P_4$.

In this step R3 encloses the four rows $P_1, \ldots, P_3$. The initial optimal coverage CV3 encloses the formal concepts $FC_3 = (\{o1, o2\}, \{A, B\})$ and $FC_4 = (\{o3\}, \{B, C\})$.

The package $P_4$ encloses only three couples: (o3,B), (o3,C) and (o3,D).

The pseudo concept containing the couple (o4, B) and (o4,C) is :

*Case 1*: the union of formal concepts ({o3,o4}, {B, C}) and ({o4},D)

*Case 2*: the formal concept ({o4}, {B, C, D}).

The computation of the relevance function induces to a negative value in the first case and positive value in the second case according to the following formula:

$$Relevance(PCF) = \left[\frac{Size(PCF)}{length(PCF) + conf(PCF)}\right] * [(length(PCF) + width(PCF) - Size(PCF)]$$

Hence, the retained formal concept is $FC_6 = (\{o4\}, \{B, C, D\})$.

Thus, the final coverage of R contains the concepts $FC_3$, $FC_4$ and $FC_6$. Finally, according to our example, we find three Item-sets : {A, B}, {B, C} and {C, D}.

## 5. Experimental study

We carried out experiments on benchmark datasets, whose characteristics are summarized in table 3. Experiments were carried out on a Pentium IV PC with a CPU clock rate of 3.06 Ghz and a main memory of 512 MB.

| Dataset | # Transactions | #Items |
|---|---|---|
| T10I4D | 1000000 | 100 |
| Mushroom | 8124 | 128 |
| T20I6D | 1000 | 9 |
| Tic Tac Toe | 958 | 10 |

TABLE 2. BENCHMARK DATASET CHARACTERISTICS

To stress on the performance of our approach, we compare our proposal to the two pionnering methods in the same trend in the litterature namely the Apriori algorithm and IAR approach. The latter is also FCA based approach.

To analyze the data, we choose the following values of parameters: MinSup=0.35 and MinConf=0.75.

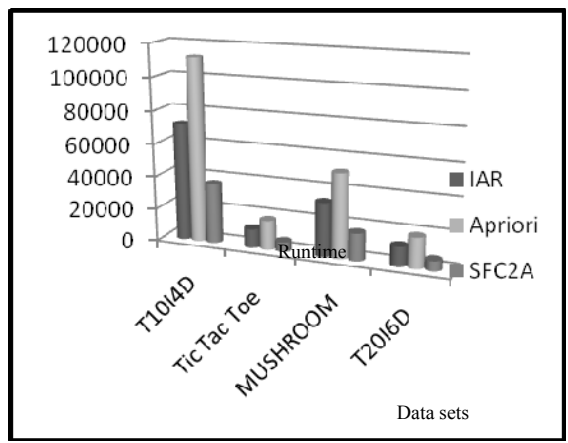

FIGURE 4. RUNTIME COMPARAISON OF SFC2A, IAR AND APRIORI ALGORITHMS

Figure 4 sketches the runtime measured in seconds for the three methods. We notice that the three algorithms keep the same behavior overall the data sets. The Apriori method takes the greater time, since it is based on an exhaustive approach to test all the possible combinations. Our method SFC2A outperforms the two algorithms IAR and Apriori thanks to its semantic selection of the frequent itemsets during the association rules extraction through the use of the relevance measure.

## 6. Conclusion

In this paper, we focused on extraction association rules based on formal concept analysis. In fact, we assume that an itemset is presented by the intent and the extent of a Formal Concept. Thus, we introduce a new approach SFC2A backboned on semantic relationship on the formal concept used for association rule extraction. The carried out experiments of our proposal showed the performance of our method compared to the pionnering approaches in the same trend.

Future works will focus on: (1) the consideration of fuzzy context, (2) the study of the extraction of ''generic association rules'' inspired from our proposal.